\documentclass[proof]{pasj00}
\begin{document}
\SetRunningHead{K. Yoshikawa et al.}{PPPM and TreePM methods on GRAPE}
\title{PPPM and TreePM Methods on GRAPE Systems \\for Cosmological N-body Simulations}

\author{Kohji \textsc{Yoshikawa}}
\affil{Department of Physics, School of Science,\\
University of Tokyo, Tokyo 133-0033}
\email{kohji@utap.phys.s.u-tokyo.ac.jp}
\and
\author{Toshiyuki \textsc{Fukushige}}
\affil{Department of General System Studies, College of Arts and
Sciences,\\ University of Tokyo, Tokyo 153-8902}
\email{fukushig@provence.c.u-tokyo.ac.jp}

\Received{}
\Accepted{}
\Published{}

\maketitle

\KeyWords{methods: n-body simulations --- cosmology: miscellaneous}

\begin{abstract}
 We present Particle--Particle--Particle--Mesh (PPPM) and Tree
 Particle--Mesh (TreePM) implementations on GRAPE-5 and GRAPE-6A
 systems, special-purpose hardware accelerators for gravitational
 many-body simulations. In our PPPM and TreePM implementations on GRAPE,
 the computational time is significantly reduced compared with the
 conventional implementations without GRAPE, especially under the strong
 particle clustering, and almost constant irrespective of the degree of
 particle clustering. We carry out the survey of two simulation
 parameters, the PM grid spacing and the opening parameter for the most
 optimal combination of force accuracy and computational speed. We also
 describe the parallelization of these implementations on a PC-GRAPE
 cluster, in which each node has one GRAPE board, and present the
 optimal configuration of simulation parameters for good parallel
 scalability.
\end{abstract}

\section{Introduction}

The Particle--Particle--Particle--Mesh (PPPM) method developed by
\citet{Hockney1981} has been recognized to be very versatile for
studying particle systems in many branches of physics. In astrophysics,
it has been widely used in a number of numerical simulations of
cosmological structure formation \citep{Efstathiou1985,
Bertschinger1991, Jing1994, Jing1998, Thomas1992, Yoshikawa2000,
Yoshikawa2001}, since it was first applied to cosmological $N$-body
simulation by \citet{Efstathiou1981}. This is because the PPPM method
has several advantages that (1) it intrinsically satisfies the periodic
boundary condition, (2) is faster under the light particle clustering,
and (3) has a smaller memory requirement than other $N$-body algorithms
such as the tree algorithm. The first one is important because, in
cosmological $N$-body simulations, periodic boundary condition is
required in order to perform realistic simulations of effectively
``infinite'' universe within a finite simulation volume.

Nowadays, however, the PPPM method in gravitational $N$-body simulations
loses its attraction due to the fact that under the strong particle
clustering, which is always realized through the gravitational
instability as the system evolves, the cost of computing the
gravitational forces is increased by a factor of ten, or even more,
depending on the clustering strength and the number of particles
adopted. In the PPPM method, the inter-particle force is split into two
components: a long-range force calculated using the Particle--Mesh (PM)
technique and a short-range one computed by directly summing up all the
contribution from nearby particles within a given cutoff radius. The
latter, Particle--Particle (PP) calculation, becomes unacceptably
expensive under the strong particle clustering, while the computational
cost of the PM calculation is independent of clustering degree of the
particles.

In order to alleviate the drawback of the PPPM method described above
while keeping its advantages, several modifications have been
proposed. In the adaptive PPPM (AP3M) method developed by
\citet{Couchman1991}, PP calculation is recursively split into further
PM and PP calculation by setting up a refined mesh at regions where the
particle clustering is strong. This method can significantly reduce the
computational cost. However, even with the AP3M method, the calculation
time increases as the particle clustering become strong by a factor of
ten or so, depending on the number of particles and the degree of
particle clustering. Another remedy for the PPPM method is to replace
the PP calculation for short-range forces by the tree method, which is
developed by \citet{Xu1995}, \citet{Bode2000}, \citet{Bagla2002}, and
\citet{Bode2003}, and is called tree particle mesh (TreePM) method. This
method reduces the expense of PP calculation under strong particle
clustering, yet requiring additional memory.

The use of a special-purposed computer for gravitational $N$-body
simulations, GRAPE (GRAvity PipE) system, provides us with a drastic
reduction of computational load of the PP calculation in the PPPM
method. For example, \citet{Brieu1995} implement the PPPM method on
GRAPE-3A system \citep{Okumura1993} for cosmological $N$-body
simulations. However, since the force shape of PP calculations in the
PPPM method is different from the $r^{-2}$ law, and GRAPE-3A can
calculate only the $r^{-2}$ gravitational force with softening, they had
to approximate the force shape by combining three $r^{-2}$ forces with
different softening lengths and different weights. Due to this
approximation, they found small but non-negligible systematic errors in
clustering statistics in a cold dark matter simulation.  Recently,
\citet{Susukita2004} implemented the PPPM method on MDGRAPE-2
\citep{Susukita2003}, an extension of MD-GRAPE (Molecular Dynamics
GRAPE) \citep{Fukushige1996} which can compute forces with arbitrary
shapes. The resulting performance of the PPPM implementation on
MDGRAPE-2 system is $\sim$10 times better than that of AP3M
implementation \citep{Couchman1991}.

In this paper, we present the PPPM and TreePM implementations on GRAPE-5
\citep{Kawai2000} and GRAPE-6A \citep{Fukushige2005} systems, both of
which have a capability to calculate the $r^{-2}$ forces multiplied by a
user-specified cutoff function, and have better performance than
MDGRAPE-2. We also develop the parallelized versions of these
implementations with Message Passing Interface (MPI) and estimate their
scalability. As we show below, these implementations on GRAPE-5 and
GRAPE-6A systems have sufficient advantage over those without GRAPE.

The outline of this paper is as follows. In \S~\ref{sec:grape}, the
architectures of GRAPE-5 and GRAPE-6A are briefly described.
\S~\ref{sec:algorithm} is devoted to brief introduction of the PPPM and
TreePM methods, their implementations on GRAPE-5 and GRAPE-6A, and their
parallelization scheme. We evaluate the accuracy of force calculated
using the PPPM and TreePM methods on GRAPE in \S~\ref{sec:force}, and
present their performance and memory requirement in
\S~\ref{sec:performance} and \S~\ref{sec:memory}, respectively. In
\S~\ref{sec:conclusion}, we summarize our conclusion and give discussion
on further improvement of our implementation on the future GRAPE
systems.

\section{GRAPE systems}\label{sec:grape}

GRAPE \citep{Sugimoto1990, Makino1998} is a special-purpose computer to
accelerate calculations of the gravitational interaction. GRAPE has
pipelines specialized for the interaction calculations, which is the
most expensive part of $N$-body simulations.  The GRAPE hardware,
connected to a standard PC or workstation (host computer), receives the
positions and masses of particles and calculates the interactions
between particles.  All other calculations are performed on the host
computer.

\begin{table}
\caption{GRAPE Summary}
\begin{center}
\begin{tabular}{lccccc}
\hline
\hline
 & \# of chips & \# of pipelines & Peak speed ($10^9$ int/s) & Accuracy \\
\hline
GRAPE-5	 & 8 & 16 & 1.28 & $\sim$0.2\% \\	
GRAPE-6A & 4 & 24 & 2.30 & single precision \\
\hline
\hline
\end{tabular}
\end{center}
\label{tab1}
\end{table}

In Table 1, we summarize features of GRAPE-5 and GRAPE-6A that we used
for implementations of the PPPM and TreePM methods in this
paper. GRAPE-6A is a single PCI card with four GRAPE-6 processor chips
\citep{Makino2003}, and is designed for PC-GRAPE cluster configuration
\citep{Fukushige2005}.  For details of GRAPE-5 and -6A, see
\citet{Kawai2000}, and \citet{Fukushige2005}, respectively.  The unit of
the peak speed is $10^9$ pair-wise interactions per second. The GRAPE-5
system is currently no longer produced, but it is still available in
several sites such as Astronomical Data Analysis Center in National
Astronomical Observatory Japan
\footnote{http://www.cc.nao.ac.jp/J/cc/muv/index\_e.html}.

These GRAPEs calculate the gravitational force with an arbitrary
cutoff function on particle $i$, given by equations:
\begin{equation}
 \label{eq:GRAPE_force}
{\vec f}_i = \sum_j {m_j {\vec r}_{ij} \over r_{{\rm s},ij}^3}
g(r_{{\rm s},ij}/\eta).
\end{equation}
Here ${\vec r}_j$ and $m_j$ are the position and the mass of particle
$j$, ${\vec r}_{ij} = {\vec r}_j - {\vec r}_i$, and $r_{{\rm s},ij}$ is
the softened distance between particles defined as $r_{{\rm s},ij}^2
\equiv r_{ij}^2 + \varepsilon^2$, where $\varepsilon$ is the softening
length. The function $g$ is an arbitrary cutoff function and $\eta$ is
the scaling length for the cutoff function. The cutoff function is
necessary for the calculation of PP force in the PPPM and TreePM
methods.  The force is calculated with limited accuracy ($\sim 0.2\%$
for pairwise) in GRAPE-5, and with single precision accuracy in
GRAPE-6A. In the rest of this paper, we show the results only for
GRAPE-6A system unless otherwise stated.

\section{Algorithm and Implementation}\label{sec:algorithm}

In this section, we briefly describe the algorithms of PPPM and TreePM
methods and their implementations on GRAPE systems. In addition, we also
present our parallelization of these implementations using MPI.

\subsection{PPPM}\label{ss:PPPM}

\begin{figure}[tbp]
 \begin{center}
  \leavevmode
  \FigureFile(14cm,14cm){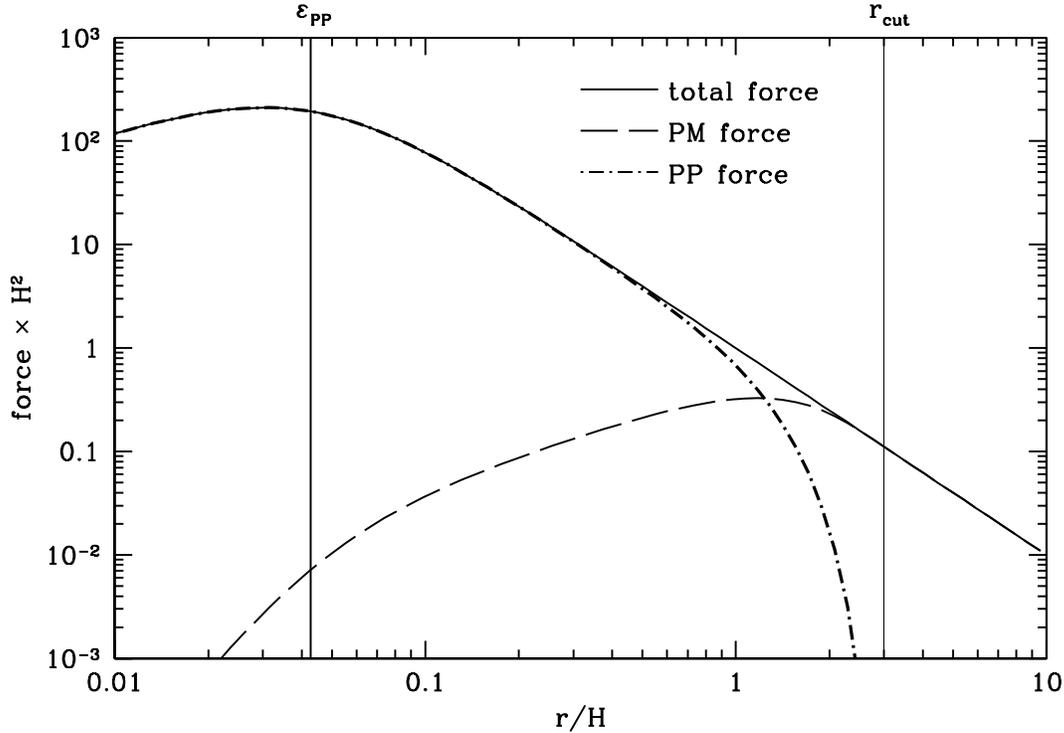}
 \end{center}
 \caption{Shapes of the total, PM and PP forces as a function of
 particle separation in units of the PM grid spacing. Vertical lines
 indicates the location of softening length $\varepsilon_{\rm PP}$ and
 the cutoff radius $r_{\rm cut}$.}\label{fig:force_shape}
\end{figure}

The PPPM method is comprised of PM and PP parts which compute long and
short-range components of inter-particle forces, respectively. In the PM
part, the mass density on a regular grid (PM grid, hereafter) is
computed by interpolating the particle positions, and the gravitational
potential is calculated by solving the Poisson equation on this grid
using Fast Fourier Transform (FFT). The force must be sufficiently
smooth so that it can be well represented by the discrete
grid. Therefore, it is assumed that each particle virtually have a
spherical density profile which is usually denoted by $S2$ profile
\citep{Eastwood1980}. In our implementation, we use the
triangular-shaped cloud (TSC) scheme to calculate the mass density field
on the PM grid, and the 4-point finite differential algorithm (FDA) to
compute the particle accelerations by interpolating the gravitational
potential field. The resulting gravitational force obeys the $r^{-2}$
law beyond a softening length $\varepsilon_{\rm PM}$ (about two or three
grid spacings), and is softened inside it. Usually, the required force
shape has smaller softening length than the grid spacing. Therefore, we
need an additional short-range component, which is non-zero only for
particle pairs with separations smaller than $\varepsilon_{\rm PM}$, to
compute the total required force. In order to compensate for this
short-range forces, the PP calculation accumulates the ``residual''
gravitational force, the required gravitational force minus that
calculated in the PM part, contributed by all nearby particles lying
within a certain cutoff radius $r_{\rm cut}$. Specifically, for a given
particle separation $r$, the force shape adopted in the PP part is given
by
\begin{equation}
 F_{\rm PP}(r)=F_{\rm tot}(r)-R(r,\varepsilon_{\rm PM}),
\end{equation}
where $F_{\rm tot}$ is the required total force, $\varepsilon_{\rm PM}$ the
softening length for the PM force, and $R(r,a)$ the force shape
calculated in the PM part given by 
\begin{equation}
 R(r,a)=\left\{
	 \begin{array}{@{\,}ll}
	  (224\xi-224\xi^3+70\xi^4  & \\
	  \hspace{1cm}+48\xi^5-21\xi^6)/35a^2 & 0 \le \xi < 1 \\
	  (12/\xi^2-224+896\xi-840\xi^2 & \\
	  \hspace{1cm}+224\xi^3+70\xi^4-48\xi^5+7\xi^6)/35a^2  & 1 \le \xi < 2 \\
	  1/r^2 & 2 \le \xi,
	 \end{array}
	\right.
\end{equation}
where $\xi=2r/a$. Note that $R(r,a)$ follows $r^{-2}$ law at $r>a$, and
is softened at $r<a$. 
As described in section \ref{sec:grape}, GRAPEs calculate the
Plummer-softened force. Therefore, the PPPM method on GRAPE systems
naturally adopts the Plummer-softened force shape for $F_{\rm
tot}(r)$ and 
\begin{equation}
 g_{\rm PP}(r,\varepsilon_{\rm PM}) = 1-R(r,\varepsilon_{\rm PM})/F_{\rm
  tot}(r),
  \label{eq:cutoff_function}
\end{equation}
for the user-specified cutoff function $g$ in
equation~(\ref{eq:GRAPE_force}). Since $R(r,\varepsilon_{\rm PM})$
deviates from $r^{-2}$ at $r<\varepsilon_{\rm PM}$, we set the cutoff
radius to $r_{\rm cut}=\varepsilon_{\rm PM}$. Throughout in this paper,
we set $\varepsilon_{\rm PM}=3H$, where $H$ is the PM grid
spacing. Figure~\ref{fig:force_shape} depicts the force shapes for the
PM and PP calculations as well as the total force shape. 

In the PP part, the simulation volume is divided into cubic chaining
cells with a side equal to or greater than $r_{\rm cut}$. For each
chaining cell, a linked-list of particles is build to keep track of all
the particles in that cell. With this linked-list, one can efficiently
calculate the contribution on each particle in the cell from all nearby
particles within the cutoff radius $r_{\rm cut}$, by accumulating the
pairwise interaction with all particles in that cell and in the adjacent
26 cells. In virtue of the symmetry of forces, we can eliminate the
unnecessary double computations of force for each particle pair. In
order to achieve this, the pairwise interactions are computed for all
pairs of particles in the current chaining cell, and for all pairs with
one particle in the current chaining cell and the other in one of 13
neighbor cells.

On GRAPE systems, however, the sum of all forces exerted on a particle
is computed for a given set of interacting particles.  Therefore, the
benefit from symmetry of forces is not available and we have to
implement the PP part in a different manner on GRAPE systems. The
overall procedures to calculate PP force using GRAPE is as follows.

\begin{itemize}
 \item[1.] The host computer calculates the PM force. 

 \item[2.]  The host computer constructs a linked-list of all particles.

 \item[3.]  Repeat 3.1 to 3.4 for all chaining cells

 \item[3.1.] The host computer sends masses and positions of particles in
	   current and 26 adjacent chaining cells to GRAPE as
	   interacting particles.

 \item[3.2.] The host computer sends positions of particles in the
	   current chaining cell to GRAPE.

 \item[3.3.] GRAPE calculates the PP forces exerted on the particles in
	   the current chaining cell, which are sent to GRAPE in 3.2.

 \item[3.4.] The PP forces computed by GRAPE in 3.3 are returned to the
	   host computer.

 \item[4.]  The host computer augments the accelerations of particles in
	   the current chaining cell using the calculated PP forces.

 \item[5.]  The host computer updates the positions and velocities of all
	   particles using the calculated PM and PP forces.

\end{itemize}
On the GRAPE-5 chip, the particle index unit optimized for the
linked-list method is equipped, with which we can reduce the
communication cost between a host computer and GRAPE-5
\citep{Kawai2000}. In stead, on the GRAPE-6A system, since the particle
index unit is not equipped, PP force calculation is done on a host
computer (not on GRAPE-6A) if number of particles in the current
chaining cell is less than a threshold $n_{\rm th}$ in order to reduce
the communication overhead between a host computer and GRAPE-6A. In this
paper, we set $n_{\rm th}=10$. Thus, the implementation of PP part for
GRAPE-5 is slightly different from that for GRAPE-6A so as to make use
of the particle index unit.

The spacing of PM grid controls the performance and memory requirement
of the implementation. If we adopt larger spacing, we have larger cutoff
radius and thus computational cost for PP part increases. To the
contrary, if smaller $H$ is adopted, the cost for PM part increases and
we needs larger memory requirement. In the conventional PPPM
implementation, $H$ is usually set to $\simeq L/(2N^{1/3})$, where $L$
is the size of simulation volume.  In order to use GRAPE system
efficiently, it is more advantageous to compute accelerations of as many
particles as the number of (virtual) pipelines on GRAPE systems
(96 for GRAPE-5 and 48 for GRAPE-6A) in the procedure 3.3
of the PP calculation listed above. In order to achieve this, in our
implementation, the spacing of the PM grid $H$ is set to 
\begin{equation}
 H=H_0\equiv L/N^{1/3},
\end{equation}
which is two times larger than that usually adopted in the conventional
PPPM implementation. Accordingly, we adopt larger $\varepsilon_{\rm PM}$
and $r_{\rm cut}$ since we set $r_{\rm cut}=\varepsilon_{\rm PM}$ and
$\varepsilon_{\rm PM}=3H$.

\subsection{TreePM}

TreePM is a method in which the direct calculation of PP force in the
PPPM method is replaced by the hierarchical tree algorithm. In this
section we briefly describe our implementation of the TreePM method on
GRAPE.

Several studies on the Tree + PM hybrid implementations were already
reported \citep{Xu1995, Bode2000, Bagla2002, Bode2003, Dubinski2003},
and they are divided into two categories. \citet{Xu1995},
\citet{Bode2000} and \citet{Bode2003} developed TPM in which tree
algorithm is used for short range force only in dense cluster
regions. Another is developed by \citet{Bagla2002} and
\citet{Dubinski2003}, in which PP force is calculated with tree
algorithm throughout the entire simulation volume.  We implement the
latter algorithm on GRAPE.
 
We modify a Barnes-Hut tree code \citep{Barnes1986}, which is originally
used with vacuum boundary on GRAPE-5 \citep{Fukushige2001}, so that (i)
the functional form of gravitational force is $1/r^2$ with a cutoff,
expressed in equation (\ref{eq:cutoff_function}), and (ii) the periodic
boundary condition is taken into account in the tree traversal.  We
incorporated Barnes' modified algorithm \citep{Barnes1990} in order to
make GRAPE work efficiently as described in \citet{Makino1991}. In the
original algorithm, tree traversal is performed for each particle. In
the modified algorithm, tree traversal is performed for a group of
neighboring particles and an interaction list is created. GRAPE
calculates the force from particles and the nodes in this interaction
list to particles in the group.

The calculation procedure of the TreePM method on GRAPE is as follows:

\begin{itemize}
\item[1.] The host computer calculates the PM force. 

\item[2.1.] The host computer constructs a tree structure, and divide
particles into groups of neighboring ones using the tree structure.

\item[2.2.] Repeat 2.3.-2.6. for all groups.

\item[2.3.] The host computer creates the interaction list for a group,
and sends the data of the particles listed up to GRAPE.

\item[2.4.] Repeat 2.5.-2.6. for all particles in a group.

\item[2.5.] The host computer sends particles to be calculated to GRAPE.

\item[2.6.] GRAPE calculates the PP forces exerted on the particles, and
then returns the result to the host computer.

\item[3.]  The host computer updates the positions and velocities of all
particles using the calculated PM and PP forces.

\end{itemize}

The modified tree algorithm reduces the calculation cost of the host
computer by roughly a factor of $n_g$, where $n_g$ is the average number
of particle in groups. On the other hand, the amount of work on GRAPE
increase as we increase $n_g$, since the interaction list becomes
longer. There is, therefore, an optimal value of $n_g$ at which the
total computing time is minimum. The optimal value of $n_g$ depends on
various factor, such as the relative speed of GRAPE and its host
computer, the opening parameter and the number of particles. For the
present GRAPE-5 and -6A, $n_g=500-1000$ is close to optimal.

\subsection{Parallel PPPM and TreePM}

\begin{figure}
 \begin{center}
  \leavevmode \FigureFile(16cm, 18cm){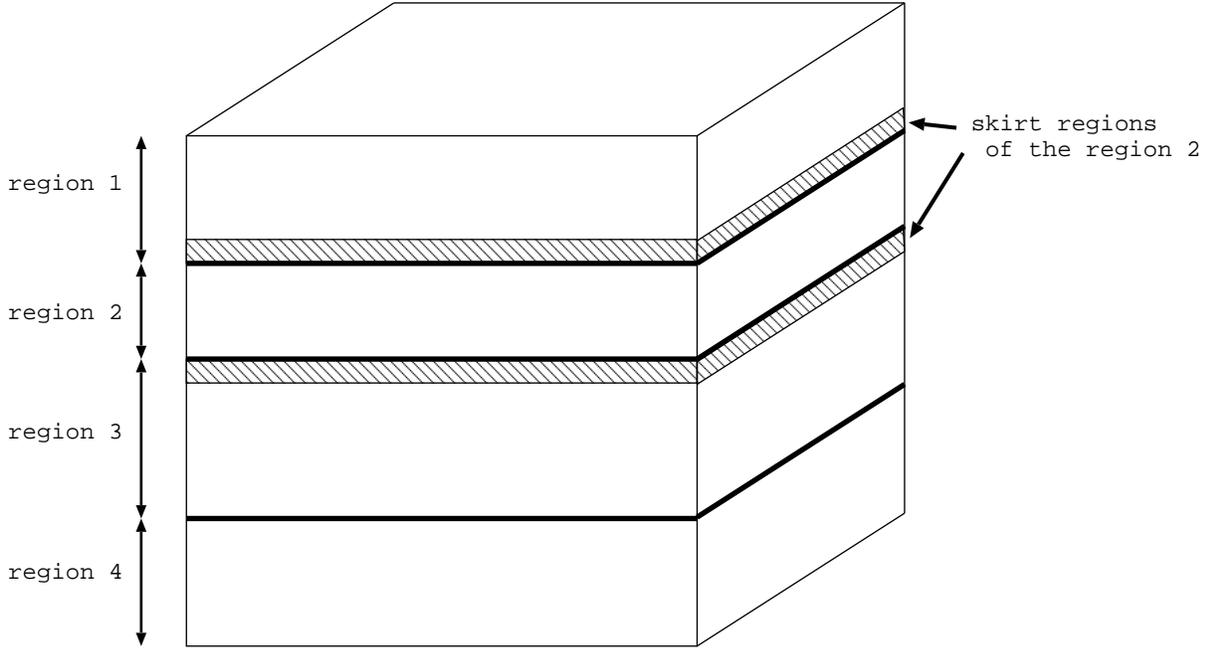}
 \end{center}
  \caption{Schematic picture of domain decomposition of simulation
  volume used in the parallel PPPM and TreePM calculations. Bold lines
  are the domain boundaries and shaded regions indicate the skirt
  regions of the region 2.\label{fig:decomposition}}
\end{figure}

We also parallelize our PPPM and TreePM implementations using the one
dimensional slice domain decomposition scheme and MPI library. Our
parallel algorithm is similar to the parallel PPPM code by
\citet{Brieu2000} and GOTPM by \citet{Dubinski2003}.  The calculation
procedure for the parallel PPPM and TreePM methods on GRAPE is as
follows:

\begin{itemize}
 \item[1.] Particles are distributed into the slice domains, and each
	   parallel node is responsible for particles in each slice
	   domain. The boundaries of the slice domains are adjusted so
	   that the number of particles in each domain is equal. As
	   these boundaries move, particles are dynamically re-assigned
	   to the parallel node responsible for the domain where they
	   reside. Figure~\ref{fig:decomposition} is a schematic picture
	   of the domain decomposition.

 \item[2.] Each node calculates the mass density on the PM grid for the
	   slice domain. The entire grid density are communicated and
	   shared among all nodes. Then, each node calculates the
	   gravitational potential on the PM grid using FFT, and
	   compute the PM forces.

 \item[3.] In the parallel PPPM, each node imports those particles from
	   the neighboring domains which reside in its skirt regions
	   with a width of $r_{\rm cut}$ from the domain boundaries. See
	   Figure~\ref{fig:decomposition} for a schematic picture of the
	   skirt regions. In each node, a linked-list is built for all
	   the particles in the slice domain and the imported particles
	   in the skirt regions. Then, the PP forces are calculated in
	   the same manner as the serial PPPM implementation.
	   
	   \bigskip

	   In the parallel TreePM, each node constructs a tree structure
	   and makes the interaction list for the neighboring
	   nodes. Each node receives the interaction lists from the
	   neighboring nodes, and reconstruct the tree structure
	   together with the received interaction lists. The PP forces
	   for particles in the slice domain are calculated using the
	   tree structure as done in the serial TreePM algorithm.

 \item[4.] The positions and velocities of particles in each node are
	   updated using the calculated PM and PP forces.

 \item[5.] Those particles which get across the domain boundaries are
	   exchanged between the neighboring nodes.

\end{itemize}

In these parallelization, the FFT is performed in a serial fashion, and
the global potential field is computed independently by each
node. Therefore, we have to communicate the density field on the PM grid
among the nodes. As can be seen in Table~\ref{tab:cputime_PPPPM} and
\ref{tab:cputime_PTreePM}, the CPU time required for these communication
is not predominant, but cannot be neglected. Therefore, the use of
parallel FFT may improve the overall parallel efficiency.

\begin{figure}
 \begin{center}
  \leavevmode
  \FigureFile(16 cm,10 cm){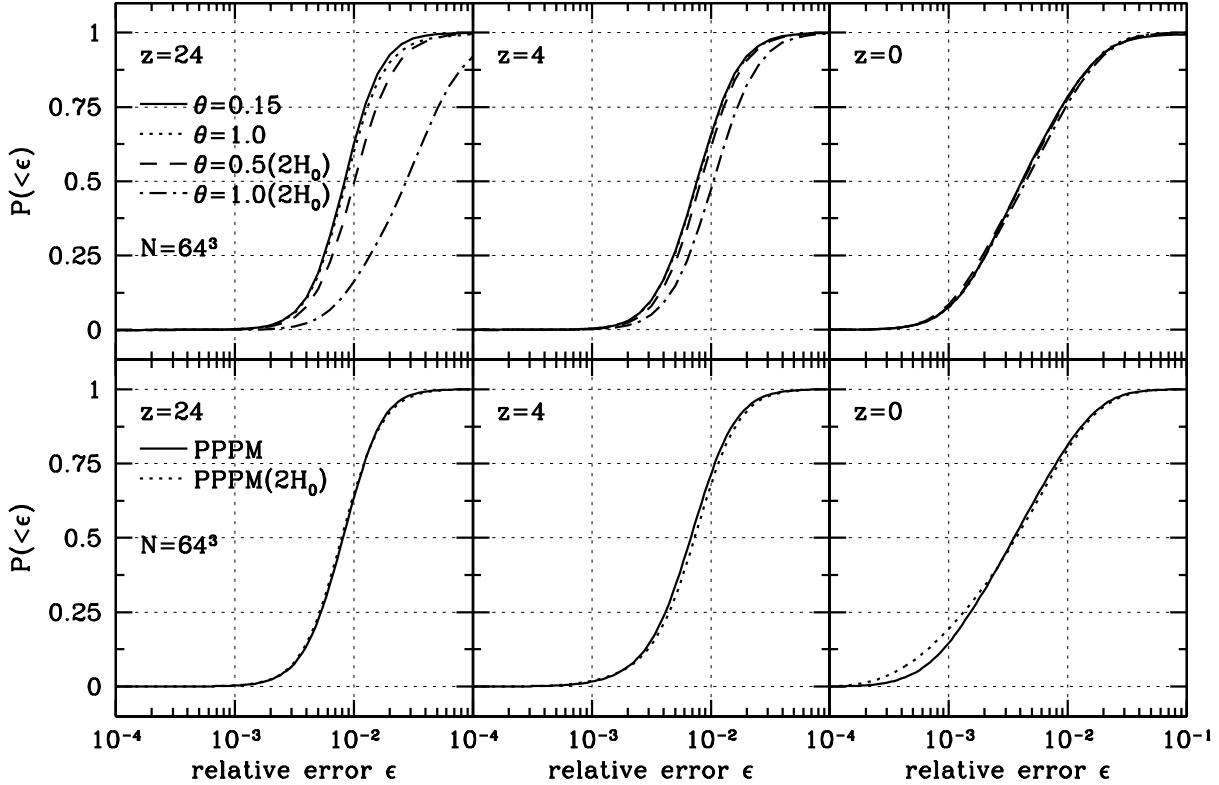}
 \end{center}
 \caption{The cumulative distribution of the relative error in the PPPM
 and TreePM methods on GRAPE for $N=64^3$. Lines labeled as ``2H$_0$''
 indicate the error distribution for the PM grid spacing of twice the
 standard value. } \label{fig:force_error_N64}
\end{figure}

\begin{figure}
 \begin{center}
  \leavevmode
  \FigureFile(16 cm,10 cm){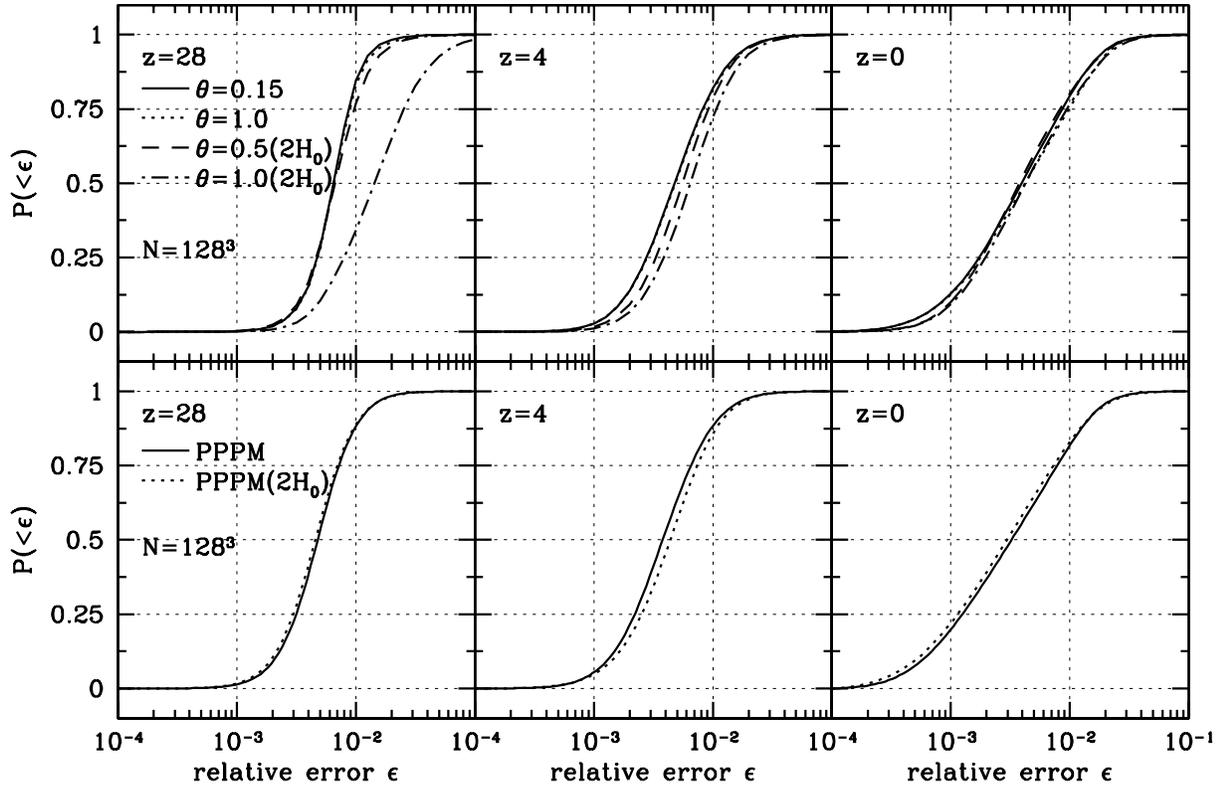}
 \end{center}
 \caption{Same as Figure~\ref{fig:force_error_N64}, but for $N=128^3$.}
 \label{fig:force_error_N128}
\end{figure}

 \begin{figure}
 \begin{center}
  \leavevmode
  \FigureFile(16 cm,16 cm){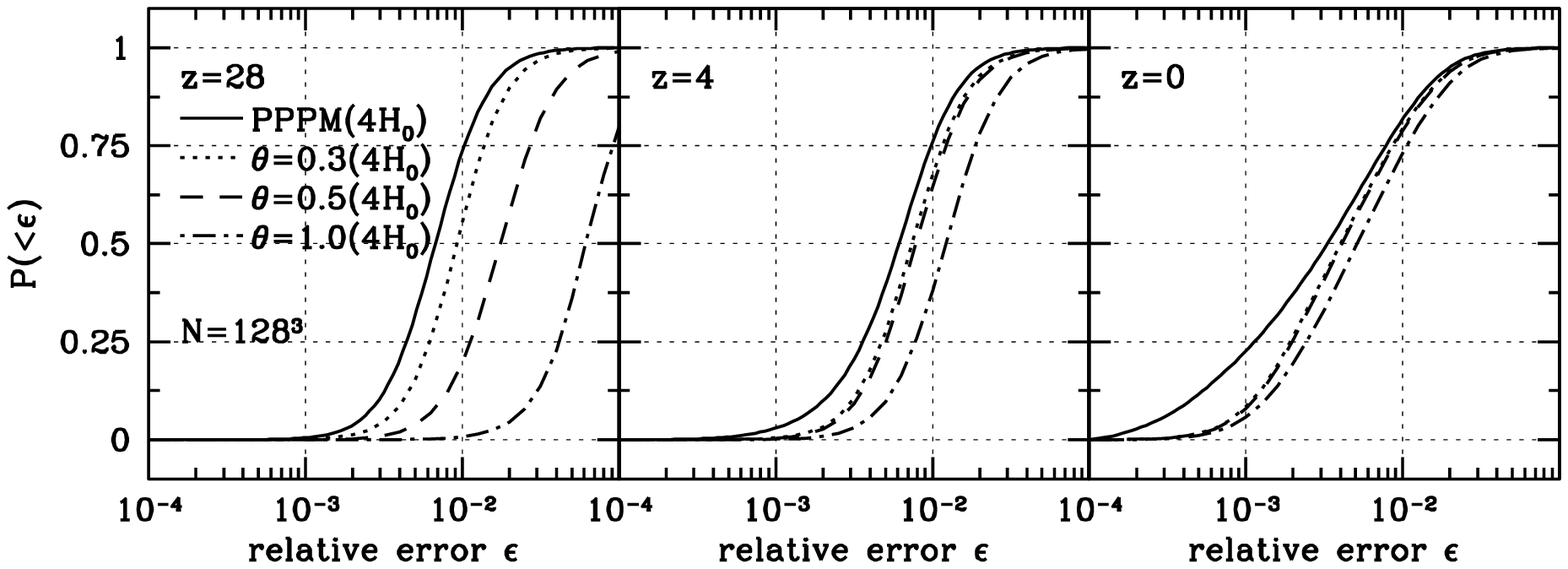}
 \end{center}
  \caption{Same as Figure \ref{fig:force_error_N128}, but for larger PM
  grid spacing, $4H_0$.}  \label{fig:force_error_4H}
 \end{figure}

\section{Force Accuracy}\label{sec:force}

Here, we discuss the force accuracy of the PPPM and TreePM methods on
GRAPE. We calculated the forces using particle distributions from dark
matter simulations in a spatially flat low density CDM (LCDM hereafter)
cosmology with $\Omega_{\rm m}=0.3$, $\Omega_\Lambda=0.7$, $h=0.7$, and
$\sigma_8=0.9$ within a box size of $L=75h^{-1}$ Mpc, where $\Omega_{\rm
m}$ is the density parameter, $\Omega_\Lambda$ the dimensionless
cosmological constants, $h$ the Hubble parameter in units of 100 km
s$^{-1}$ Mpc$^{-1}$, and $\sigma_8$ the rms density fluctuation at a
scale of 8$h^{-1}$Mpc. In this section, we only show the results
obtained using the GRAPE-6A board. The exact forces are computed using
the direct Ewald summation (whose convergence parameters are set so that
the relative force error is less than $10^{-5}$). Here, we measured the
relative force error given by
\begin{equation}
 \epsilon_i = \frac{|{\vec f_i}-{\vec f_{{\rm ew},i}}|}{|{\vec f_i}|},
\end{equation}
where ${\vec f_i}$ and ${\vec f_{{\rm ew},i}}$ are force vectors of the
$i$-th particle calculated using the PPPM or TreePM methods and the
direct Ewald summation, respectively. Figure \ref{fig:force_error_N64}
and \ref{fig:force_error_N128} show the cumulative distributions of the
relative force error $\epsilon$ in the PPPM (lower) and TreePM (upper)
methods for the number of particles of $N=64^3$ and $N=128^3$,
respectively. For the TreePM method, we show the error distributions for
various BH opening parameters, $\theta$.

As a matter of course, the force accuracy in the PPPM method is better
than that of the TreePM method, since, in the latter, the direct sum
calculation of PP part in the PPPM method is replaced by the tree
method. The force accuracy and computational cost of the tree method is
controlled by the opening parameter $\theta$. When the PM grid spacing
are set to be the standard value, $H=H_0$, we can see in these figures
that the error distributions in the TreePM method with $\theta=1.0$ and
with $\theta = 0.15$ are in good agreement with those in the PPPM
method, independent of the number of particle, $N$, and the redshift $z$
(i.e. clustering).  This means that the error due to the tree force
calculation when $\theta \le 1$ is small compared with the error in the
PM force calculation, which is nearly equivalent to the error
distribution in the PPPM method. From these results, we conclude that in
the TreePM method, the opening parameter, $\theta=1.0$, is small enough
when $H=H_0$.  In general, the force accuracy in cosmological
simulations using the tree algorithm is rather worse especially when the
particle distribution is close to homogeneous. Since the tree method
always needs to sum up the forces from all the particles or cells, and,
under the homogeneous particle distribution, forces from individual
particles or cells are much larger than the net forces, even if the
relative errors of forces contributed from each particle or cell are
small, the sum of large individual forces could results in large errors
of the net force. However, in the case of TreePM, the force error is not
so large as shown in Figure \ref{fig:force_error_N64} and
\ref{fig:force_error_N128}, simply because the long range force is
computed using the PM method, and the contribution of the tree PP force
is small.

We also check the force accuracy when the PM grid spacing are larger
than the standard value, $H_0$. Using larger PM grid spacing, memory space
for the PM calculations can be saved and a better scalability is
expected for our present parallel implementation as will be discussed in
the next section. In figure \ref{fig:force_error_N64} and
\ref{fig:force_error_N128}, the cumulative distributions of the relative
force error for twice the standard value, $2H_0$, are also shown. In the
PPPM method, the error distributions are almost unchanged compared with
the case of the standard PM grid spacing. However, in the TreePM
method, the force accuracy becomes worse when the particle distribution
are close to homogeneous, since the contribution of PP calculation to
the overall force computation is larger. At the beginning of simulations
where the particle distribution is close to homogeneous, force
calculation with $\theta \le 0.5$ is required. As the clustering
proceeds the error due to tree algorithm becomes smaller, and $\theta =
1.0$ is small enough at $z\le 4$. Figure \ref{fig:force_error_4H} shows
the cumulative distribution for the PM grid spacing of $4H_0$. In this
case, the force error becomes much worse in the TreePM method. The force
calculations with $\theta=0.3$, $0.5$ and $1.0$ seem to be marginally
acceptable at the initial epoch, $z=4$, and $z=0$, respectively. Also in
the PPPM method, the force accuracy gets worse, but only slightly.

\section{Performance}\label{sec:performance}

In this section, we present systematic measurements of the performance
of the PPPM and TreePM implementations on GRAPE systems. We use dark
matter simulations in LCDM cosmology within a box size of $L=75h^{-1}$
Mpc.

\subsection{Serial Implementation}

First, we discuss the performance for a single node configuration
composed of one GRAPE-5/GRAPE-6A board and one host computer. In the
following, the host computer for GRAPE-5 is equipped with Intel Pentium
4 (2.4GHz E7205) with 2GB of PC2001 memories, and that for GRAPE-6A has
Intel Pentium 4 (2.8GHz, i865G) with 2GB of PC3200 memories. We use gcc
compiler (version 3.2.2) on Redhat 9.0 (Linux kernel 2.4.20-8)
throughout in these measurement. We adopt the PM grid spacing of $H=H_0$
in the simulations with GRAPE and $H=H_0/2$ in those without GRAPE
unless otherwise stated.

\begin{figure}
 \begin{center}
  \leavevmode \FigureFile(12cm,10cm){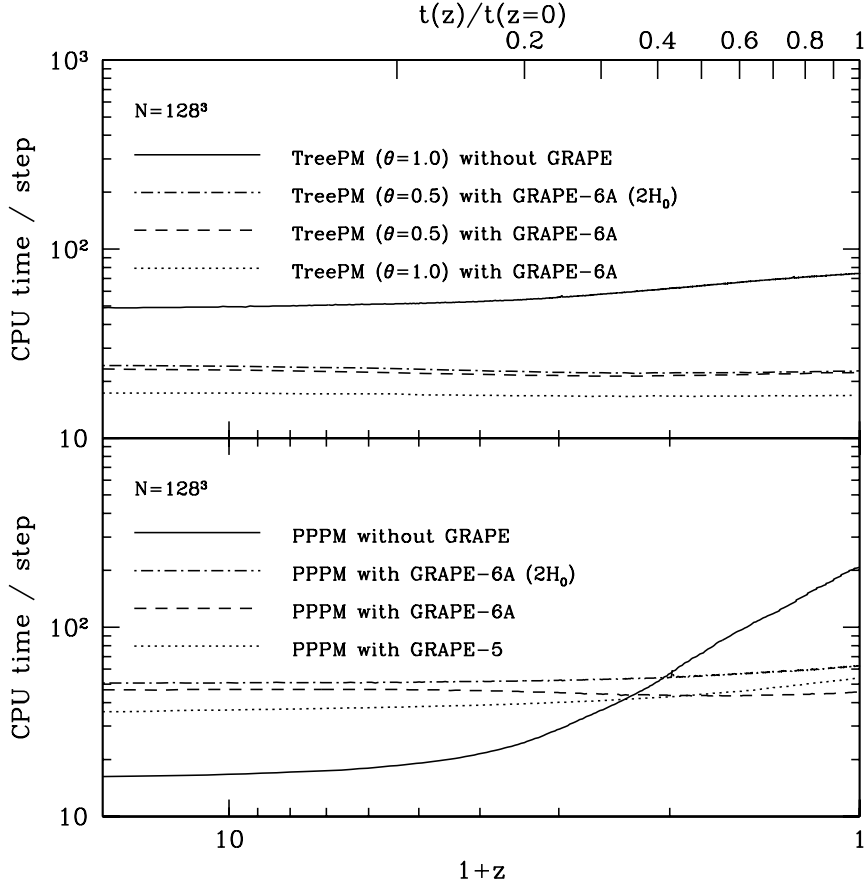} 
 \end{center}
\caption{The CPU time per step for the PPPM and TreePM methods. Lines
  labeled as ``2H$_0$'' indicate the results for the PM grid spacing of
  twice the standard value. \label{fig:cputime_PPPM}}
\end{figure}

\begin{figure}
 \begin{center}
  \leavevmode
  \FigureFile(12 cm,10 cm){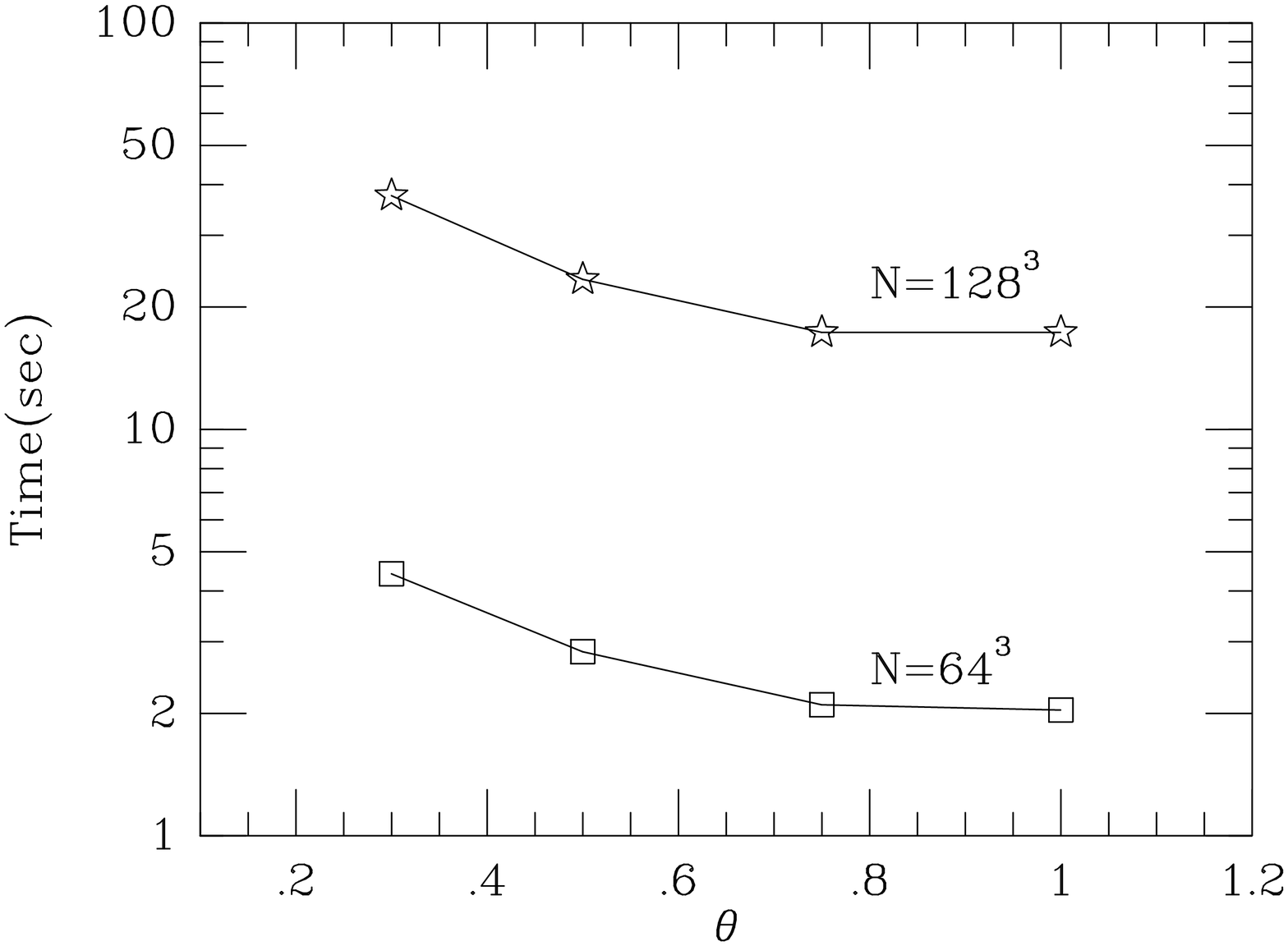}
 \end{center}
 \caption{The CPU time per timestep for the TreePM method on GRAPE with
 single node, as a function of the opening parameter, $\theta$}
 \label{fig:TreePM_opening_parameter}
\end{figure}

Figure~\ref{fig:cputime_PPPM} shows the CPU time per timestep as a
function of redshift using the PPPM (lower) and TreePM (upper) methods
for number of particles of $N=128^3$. In both panels, we show the CPU
time per timestep without GRAPE just for comparison. For the runs
without GRAPE, the PPPM code is based on the implementation adopted in
\citet{Jing1998}, and the TreePM code is modified so that the average
number of particles in groups $n_g$, for which the tree traversal is
performed, is reduced to $n_g=30$. In the upper horizontal label, we
show the ticks in the cosmic time normalized to a redshift of $z=0$. In
PPPM method without GRAPE, the CPU time at $z=0$ is $\sim 13$ times
longer than that at $z=28$. This is well known drawback of the PPPM
method under strong particle clustering at a lower redshift, which leads
to the rapid increase of computational cost for PP calculation. For the
PPPM method with the standard PM grid spacing $H$, although the CPU time
with GRAPE-5 or GRAPE-6A at a higher redshift is 2--3 times longer than
that without GRAPE, the use of GRAPE significantly reduces the CPU time
at a lower redshift $z<1.5$ or $t(z)/t(z=0)>0.3$, where $t(z)$ is the
cosmic time at a redshift of $z$. For the TreePM method with the
standard PM grid spacing $H_0$, the CPU time on GRAPE is less than half
of that for the PPPM method, if we adopt $\theta>0.5$. In comparison
with the TreePM method without GRAPE, when $\theta=1.0$ is adopted, we
can obtain a speedup factor of 2.8 and 4.4 with GRAPE at redshift of
$z=28$ and $z=0.0$, respectively. This is in virtue of the tree
algorithm used in the PP part instead of direct force calculation
adopted in the PPPM method. The most remarkable feature of the PPPM and
TreePM methods on GRAPE is that the CPU time per timestep is almost
constant irrespective of redshift, or equivalently, the strength of
particle clustering. In Figure~\ref{fig:cputime_PPPM}, we also show the
CPU time per timestep when we adopt larger PM grid spacing. Adopting
larger PM grid spacing in the PPPM method increases the CPU time per
timestep due to the increase of PP calculation. On the other hand, for
the TreePM method, the CPU time is not much affected even if we adopt
larger PM grid spacing. Figure \ref{fig:TreePM_opening_parameter} shows
the CPU time per timestep plotted versus the opening parameters of tree
algorithm for $N=64^3$ and $128^3$.  For $\theta>0.7$, the CPU time per
timestep is almost constant, and for $\theta\simeq0.3$, it is close to
that of the PPPM method with the same number of particles.

Tables \ref{tab:cputime_PPPM} and \ref{tab:cputime_TreePM} give a
breakdown of the calculation time for one timestep in the PPPM method
and TreePM method, respectively. In these tables, "others", "tree", and
"comm" indicate the time for mesh assignment and calculation of
potentials and forces in the PM calculation, the time for tree
construction and traversal, and the time for data communication between
host computer and GRAPE and data conversion on host computer,
respectively. Times labeled as ``GRAPE'' indicate those consumed by
GRAPE chip, which is estimated by the number of interactions calculated
by GRAPE chip. Specifically, these times are estimated using following
formula:
\begin{equation}
 T=(N_{\rm int}+100N)/P,
\end{equation}
where $N_{\rm int}$, $N$ and $P$ are the number of interactions and
particles, and computational speed of the GRAPE chip, respectively. $P$
is $1.28\times10^9$ (interactions/sec) and $2.3\times10^9$
(interactions/sec) for GRAPE-5 and GRAPE-6A, respectively. The second
term in the parenthesis indicates the pipeline delay in the GRAPE
chip. In table~\ref{tab:cputime_PPPM}, the significant differences in
the PM calculation time for the runs with and without GRAPE arise from
the difference of the adopted PM grid spacing. We can see that the most
dominant part is the communication between GRAPE and a host computer,
which will be improved in the new versions of GRAPE systems.

\begin{table}
 \caption{The calculation time with the serial PPPM method for one
 timestep with GRAPE-5 and GRAPE-6A, and without
 GRAPE\label{tab:cputime_PPPM}}
 \begin{center}
  \begin{tabular}{l|rrrrrrrr}
   \hline
   \hline
   $N$ & $64^3$ & $64^3$ & $128^3$ & $128^3$ & $128^3$ & $128^3$ & $128^3$ & $128^3$ \\
   redshift & 24 & 24 & 28 & 28 & 28 & 0 & 0 & 0 \\
   GRAPE & 5 & 6A & none & 5 & 6A & none & 5 & 6A \\
   \hline
   PM(sec/step) & 0.48 & 0.4  & 9.42 & 4.07 & 3.43 & 8.36 & 4.19 & 3.38 \\
   \quad FFT    & 0.02 & 0.02 & 2.99 & 0.25 & 0.21 & 2.98 & 0.25 & 0.22 \\
   \quad others & 0.46 & 0.38 & 6.43 & 3.82 & 3.22 & 5.38 & 3.94 & 3.27 \\
   \hline
   PP(sec/step) & 4.97 & 5.42 & 6.37 & 30.8 & 42.5 & 200 & 49.5 & 39.1 \\
   \quad GRAPE  & 0.48 & 0.25 & N/A  & 1.73 & 0.90 & N/A & 11.2 & 5.81 \\
   \quad comm   & 4.49 & 5.17 & N/A  & 30.5 & 41.6 & N/A & 38.3 & 33.3 \\
   \hline
   Total(sec/step) & 5.43 & 5.84 & 16.1 & 35.3 & 45.9 & 210 & 54.0 & 42.5 \\
   \hline
  \end{tabular}
 \end{center}
\end{table}

\begin{table}
 \caption{The calculation time with the serial TreePM method for one
 timestep with $\theta=0.5$ and $1.0$}
 \begin{center}
  \begin{tabular}{l|rrrrrrrr}
   \hline
   \hline
   $N$         & $64^3$  & $64^3$ & $128^3$ & $128^3$ & $128^3$ & $128^3$ & $128^3$ &
   $128^3$ \\
   $\theta$    & $1.0$   & $0.5$  & $1.0$ & $1.0$ & $0.5$ & $1.0$ & $1.0$   & $0.5$   \\ 
   redshift    & $24$    & $24$   & $28$ & $28$ & $28$ & $0$ & $0$ & $0$
   \\
   GRAPE       & 6A   & 6A   & none & 6A & 6A & none & 6A & 6A \\
   \hline
   PM(sec/step)& 0.41 & 0.42 & 8.80 & 3.86 & 3.73 & 8.40 & 3.75 & 3.69 \\
   \quad FFT   & 0.02 & 0.02 & 2.99 & 0.20 & 0.20 & 2.99 & 0.20 & 0.20 \\
   \quad others& 0.40 & 0.41 & 5.81 & 3.76 & 3.63 & 5.41 & 3.65 & 3.59 \\
   \hline
   PP(sec/step)& 1.58 & 2.37 & 39.81 & 13.13 & 19.33 & 66.41 & 13.07 & 18.26 \\
   \quad tree  & 0.27 & 0.47 & 17.01 &  2.13 & 3.62  & 21.61 & 2.91 &  4.68 \\
   \quad comm  & 1.10 & 1.53 & N/A & 9.31 & 12.62   & N/A & 8.32   & 10.66 \\
   \quad GRAPE\qquad & 0.21 & 0.37& N/A & 1.69   & 3.09 & N/A   & 1.84   &  2.92 \\
   \hline
   Total(sec/step)& 2.03 & 2.83 & 48.61 & 17.32   & 23.39   & 74.78 & 16.88  & 22.28 \\
   \hline
  \end{tabular}
 \end{center}
 \label{tab:cputime_TreePM}
\end{table}

\subsection{Parallel Implementation}

\begin{figure}
 \begin{center}
  \leavevmode
  \FigureFile(15 cm,10 cm){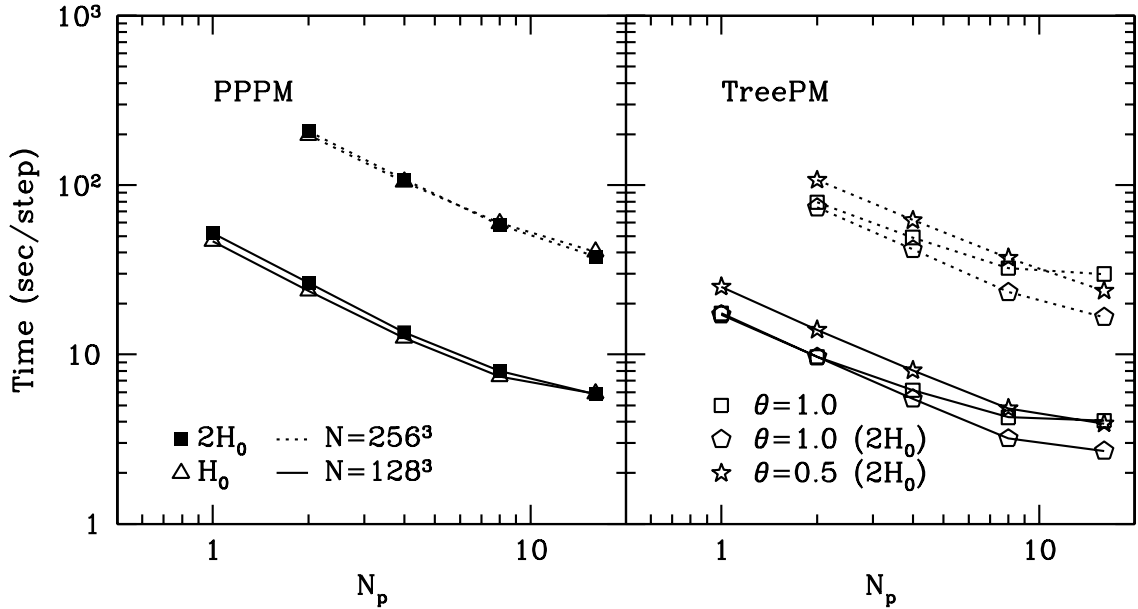}
 \end{center}
\caption{The CPU time per timestep for parallel implementation of the
TreePM method on GRAPE as a function of the number of node, $N_{\rm
p}$.}  \label{fig:scalability}
\end{figure}

Next, we discuss the parallel performance.  Figure \ref{fig:scalability}
shows the CPU time (measured with {\tt MPI\_wtime()}) for the parallel
PPPM (left) and TreePM (right) methods with GRAPE as a function of the
number of nodes $N_{\rm p}$ for $N=128^3$ and $256^3$. We could not
measure that for $N_{\rm p}=1$ and $N=256^3$ owing to a short of
memory. The parallel system consists of at maximum 16 nodes each of
which consists of one host computer and one GRAPE-6A board. The host
computers have Intel Pentium 4 (2.8CGHz, i865G) with 2GB of PC3200
memories for the 1st to 8th node, and Intel Pentium 4 (2.4BGHz, E7205)
with 2GB of PC2100 memories for the 9th to 16th node. We use gcc
compiler (version 3.2.2) and MPICH (version 1.2.5.2) on Redhat 9.0. All
nodes are connected with Gbit Ethernet via a single switch. We plot the
median of those for the first 9 steps from the initial conditions
($z=28$ for $N=128^3$ and $z=39$ for $N=256^3$).

The right panel of figure~\ref{fig:scalability} shows that our present
implementation of the parallel TreePM method obtains moderately good
performance up to $N_{\rm p}=8$.  In the case of $\theta=1.0$ in which
the PP part is less expensive than the $\theta=0.5$ case, the parallel
scalability is not so good because the time for communication of the PM
grid data logarithmically increases as $N_{\rm p}$ increases in our
present implementation.  This can be seen in Table
\ref{tab:cputime_PTreePM}, which gives the breakdown of the calculation
time per timestep of the 1st node. In the table, "others" include the
time for the date exchange and the synchronization. We can see that the
scalability is improved in the case of the PM grid spacing,
$2H_0$. Adopting larger PM grid spacing reduces the communication cost
for the PM grid data and the scalability is improved, despite that
smaller opening parameter $\theta$ is required for reasonable force
accuracy at initial homogeneous phase as discussed in the previous
section.  In the parallel PPPM method, the scalability is better than
the TreePM method, since the computational cost of PP part is
predominant compared with that of the PM part. On the contrary to the
parallel TreePM method, larger PM grid spacing does not improve the
total performance and the parallel efficiency. As can be seen in Table
\ref{tab:cputime_PPPPM}, this is because the larger cutoff radius
$r_{\rm cut}$ leads to the increase in computational cost required in
the PP calculation, despite of the reduction of the communication
overhead in the PM part. It must be noted that, for both of the parallel
PPPM and TreePM methods, the communication overhead between GRAPE and
host computers is dominant time sink, as is also the case with the
serial implementations.

\begin{figure}
 \begin{center}
  \leavevmode
  \FigureFile(15 cm,10 cm){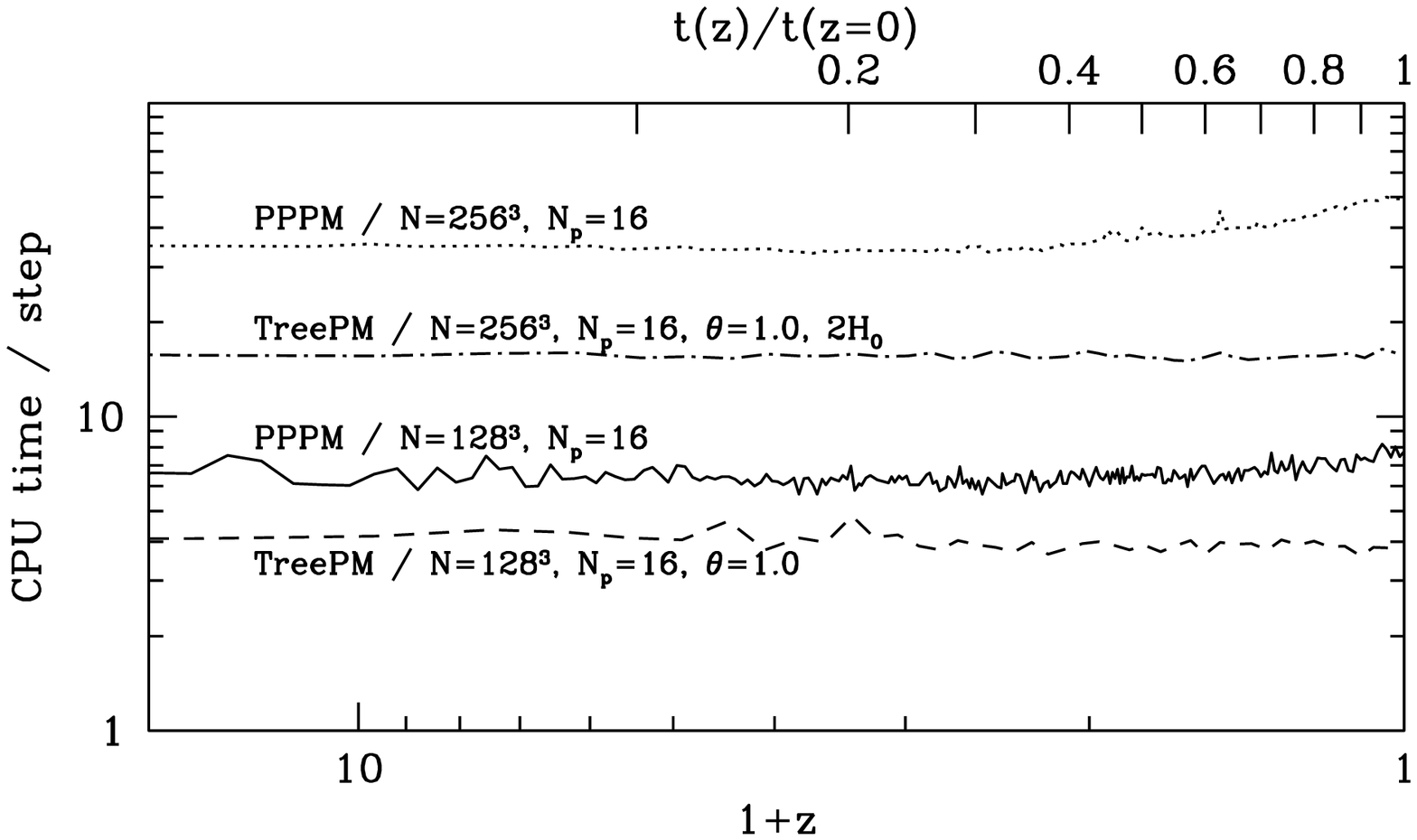}  
 \end{center}
 \caption{The CPU time per step for the parallel PPPM and TreePM runs
 with 16 nodes. The upper horizontal axis shows the cosmic time
 normalized to $z=0$. Lines labeled as ``2H$_0$'' indicate the results
 for the PM grid spacing of twice the standard
 value. }\label{fig:parallel_timing}
\end{figure}

In figure~\ref{fig:parallel_timing}, we plot the CPU time per timestep
as a function of redshift for $N=128^3$ and $N=256^3$ LCDM simulations
using the parallel PPPM and TreePM methods on 16 nodes. Similar to the
serial implementations, the CPU time is almost constant irrespective of
the redshift or particle clustering. The CPU time in the parallel PPPM
method for $N=256^3$ slightly increases at $1+z\ltsim 1.5$, and the
similar feature can be vaguely seen for the parallel PPPM run with
$N=128^3$. This is due to the difference in the number of particles in
the skirt regions for each node, which leads to the imbalance of the PP
calculation cost among the parallel nodes. In order to prevent this
feature, we have to consider more sophisticated schemes for the domain
decomposition.

\begin{table}
 \caption{Breakdown of the calculation time per timestep for the
 parallel PPPM method on GRAPE-6A.\label{tab:cputime_PPPPM}}
 \begin{center}
  \begin{tabular}{l|rrrr|rrrr}
   \hline\hline
   $N$ & $128^3$ & $128^3$ & $128^3$ & $128^3$ & $256^3$ & $256^3$ &
   $256^3$ & $256^3$ \\
   PM grid spacing & $H_0$ & $H_0$ & $2H_0$ & $2H_0$ & $H_0$ & $H_0$ & $2H_0$ & $2H_0$\\
   $N_{\rm p}$ & 2 & 8 & 2 & 8 & 2 & 8 & 2 & 8 \\
   \hline
   PM (sec/step)& 2.26 & 1.78 & 1.41 & 0.49 & 22.36 & 14.31 & 12.60 & 4.25\\
   \quad internode comm & 0.31 & 0.95 & 0.08 & 0.11 & 2.26 & 7.20 & 0.31
   & 0.91\\
   \quad FFT & 0.20 & 0.21 & 0.02 & 0.02 & 2.64 & 2.71 & 0.21 & 0.20 \\
   \quad others & 1.75 & 0.62 & 1.31 & 0.36 & 17.46 & 4.40 & 12.08 & 3.14 \\
   \hline
   PP (sec/step)        & 21.40 & 5.27 & 24.94 & 7.23 & 174.32 & 44.04 &
   194.85 & 53.82 \\
   \quad internode comm & 0.11 & 0.12 & 0.18 & 0.31 & 0.41 & 0.39 & 0.81
   & 0.89\\
   \quad comm           & 20.8 & 5.03 & 21.48 & 5.99 & 170.80 & 42.86 &
   171.43 & 46.54 \\
   \quad GRAPE          & 0.46 & 0.12 & 3.28 & 0.93 & 3.11 & 0.79 & 22.61
   & 6.19 \\
   \hline
   others (sec/step)    & 0.15 & 0.26 & 0.27 & 0.26 & 0.83 & 1.08 & 0.73
   & 0.81\\
   \hline
   Total (sec/step)  & 23.81 & 7.31 & 26.62 & 7.98 & 197.51 & 59.43 & 208.18 & 58.88\\
   \hline
  \end{tabular}
 \end{center}
\end{table}

\begin{table}
 \caption{Same as Table~\ref{tab:cputime_PPPPM}, but for the parallel
 TreePM method.}
 \begin{center}
  \begin{tabular}{l|rrrrrr|rrrrrr}
   \hline
   \hline
$N$                   &$128^3$&$128^3$&$128^3$&$128^3$&$128^3$& $128^3$ 
& $256^3$&$256^3$&$256^3$&$256^3$&$256^3$& $256^3$ \\
$\theta$              & $1.0$ & $1.0$ & $1.0$ & $1.0$ & $0.5$ & $0.5$
& $1.0$ & $1.0$ & $1.0$ & $1.0$ & $0.5$ & $0.5$ \\ 
PM grid spacing       & $H_0$   & $H_0$   & $2H_0$  & $2H_0$  & $2H_0$  & $2H_0$
& $H_0$   & $H_0$   & $2H_0$  & $2H_0$  & $2H_0$  & $2H_0$   \\
$N_{\rm p}$             & $2$   & $8$   & $2$   & $8$   & $2$   & $8$
& $2$   & $8$   & $2$   & $8$   & $2$   & $8$   \\
\hline
PM(sec/step)          & 2.35  & 1.58  & 1.53  & 0.52  & 1.49  & 0.53
& 22.26 & 14.35 & 13.07 & 4.28  & 13.07 & 4.32 \\
\quad internode comm  & 0.33  & 0.89  & 0.07  & 0.13  & 0.04  & 0.14
& 2.40  &  7.12 &  0.32 & 0.89  &  0.32 & 0.94 \\
\quad FFT             & 0.20  & 0.20  & 0.02  & 0.02  & 0.02  & 0.02 
& 2.64  &  2.74 &  0.20 & 0.20  &  0.20 & 0.20 \\
\quad others          & 1.82  & 0.49  & 1.44  & 0.37  & 1.43  & 0.37
& 17.22 &  4.49 & 12.55 & 3.19  & 12.55 & 3.18 \\ 
\hline
PP(sec/step)          & 7.08  & 2.45  & 7.90  & 2.57  &12.18  & 4.19 
& 54.95 & 16.83 & 58.63 & 17.65 & 92.06 & 31.87 \\
\quad internode comm  & 0.11  & 0.43  & 0.17  & 0.38  & 0.17  & 0.60 
&  0.72 &  2.54 &  0.91 &  2.64 &  0.95 &  2.27 \\
\quad tree  & 1.17  & 0.44 & 1.25  & 0.39 & 2.11 & 1.02 & 7.57 & 3.06 & 10.49 & 3.11 & 18.09 & 4.96 \\
\quad comm  & 4.96  & 1.36 & 5.53  & 1.56 & 8.13 & 2.12 &39.82 & 9.50 & 39.48 & 9.95 & 58.55 & 20.98 \\
\quad GRAPE & 0.84  & 0.22 & 0.95  & 0.24 & 1.77 & 0.45 & 6.84 & 1.73 & 7.75  & 1.95 & 14.47 & 3.66 \\
\hline
others(sec/step)      & 0.25  & 0.23  & 0.24  & 0.10  & 0.30  & 0.30
&  1.93 & 1.16  & 2.00  & 1.40  & 1.95  & 0.97  \\
\hline
Total(sec/step)       & 9.68  & 4.26  & 9.67  & 3.19  &13.97  & 4.79
& 79.14 & 32.34 & 73.71 & 23.33 & 107.08& 37.11 \\
\hline
  \end{tabular}
 \end{center}
 \label{tab:cputime_PTreePM}
\end{table}

\begin{figure}[tbp]
 \begin{center}
  \leavevmode \FigureFile(15 cm,10 cm){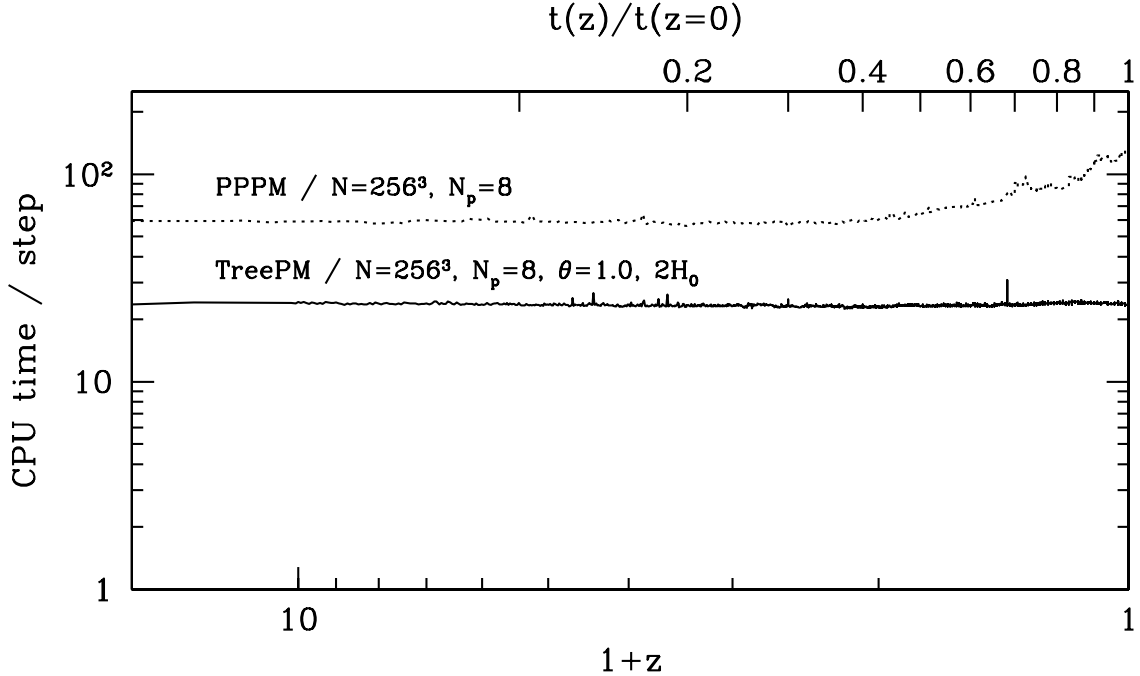} 
 \end{center}
 \caption{The
 CPU time per step for the parallel runs for the initial condition of
 the Santa Barbara Cluster Comparison Project. }\label{fig:SB_timing}
\end{figure}

So far, we present the parallel performance for the runs in a relatively
large cosmological volume. It is also interesting to measure the
parallel performance for very strongly clustered systems, for which it
might be expected that the simple one dimensional domain decomposition
technique is not so effective. As such an example, we perform parallel
runs for the initial condition of the Santa Barbara Cluster Comparison
Project \citep{Frenk1999}, which is a constrained initial condition and
is set up so that a very massive dark halo with $\simeq 10^{15}
M_{\odot}$ is eventually formed at $z=0$. In the initial condition,
Einstein de-Sitter CDM cosmology with $\Omega_0=1$, $\Omega_\Lambda=0$,
$h=0.5$ and $\sigma_8=0.9$ is assumed and the size of the simulation
volume is set to $L=64h^{-1}$Mpc. Figure~\ref{fig:SB_timing} shows the
CPU time per step for the parallel runs with 8 PC-GRAPE nodes for both
of the parallel PPPM and TreePM implementations. In the parallel TreePM
implementation, we adopt the opening parameter of $\theta=1.0$ and the
PM grid spacing of $H=2H_0$.  In the parallel TreePM run, the CPU time
per step is almost constant and the simple one dimensional domain
decomposition works well even under the strong clustering at lower
redshift. On the other hand, in the parallel PPPM run, significant
increase of CPU time per step can be seen at lower redshift
$1+z\ltsim1.5$, which is again caused by the imbalance of number of
particles in the skirt regions. Since, at lower redshift, we have
strongly clustered particles as a very massive dark halo, such an
imbalance of number of particles in the skirt regions is more serious
than that for unconstrained runs shown in
figure~\ref{fig:parallel_timing}.

\section{Memory Consumption}\label{sec:memory}

In the PPPM implementation, the memory consumption can be approximately
estimated as
\begin{equation}
 112\left({N\over128^3}\right) + 9\left(N_{\rm PM}\over128\right)^3 {\rm
  Mbytes},
  \label{eq:memory_PPPM}
\end{equation}
where $N$ is the number of particles, $N_{\rm PM}$ the number of PM grid
per side. The first term is for the data of particles and the
linked-list structure, and the second one is for the PM grid and the
data for the Green's function used in the PM calculation. Note that, for
the run with the number of particles of $N$ on GRAPE, we usually adopt
$N_{\rm PM}=N^{1/3}$ as an optimal value for the efficient use of
GRAPE. In the TreePM implementation, the memory requirement is
approximately estimated as
\begin{equation}
 \displaystyle
  104\left({N\over128^3}\right) + 60\left({N\over 128^3}\right) +
  9\left({N_{\rm PM} \over 128}\right)^3 {\rm Mbytes}, 
  \label{eq:memory_TreePM}
\end{equation}
where the first, second and third terms are for particles, tree
structure and PM grid data, respectively. $N_{\rm PM}$ is set to $N_{\rm
PM}=N^{1/3}$ as the PPPM implementation just for the efficient use of
GRAPE.  The factor 60 for tree structure is rather small compare to
usual implementations of the tree algorithm.  This is because we allow
leaf nodes to have more than one particle, and stop the subdivision of
tree node if the number of particles belonging the node is less than 30,
up to which the calculation cost does not increase so much. The memory
consumption of our PPPM implementation considerably less than that of
the TreePM method since the tree structure requires significant amount
of memory, though the memory consumption for the chaining cells used in
the PPPM method is negligible compared with the total memory
consumption.

In the parallel implementation, each node is responsible for roughly
$N/N_{\rm p}$ particles and requires an extra memory for the
communication with the neighboring nodes. In the parallel PPPM and
TreePM method, the extra memory is used for the imported particles in
the skirt regions as depicted in figure~\ref{fig:decomposition}, and for
the interaction list of the neighboring nodes, respectively.  Thus, for
the parallel PPPM method, the memory consumption in each node is
approximately expressed by replacing the number of particles $N$ in
equation (\ref{eq:memory_PPPM}) with $N/N_{\rm p}+2N/N_{\rm cc}$. The
second term roughly indicates the number of particles in the skirt
regions. Here, $N_{\rm cc}$ is the number of chaining cells per side and
is usually set to $N_{\rm cc}\simeq L/r_{\rm cut}=N^{1/3}/3$ in our
implementation.  For the parallel TreePM method, we can approximate the
memory consumption for each node by replacing $N$ in equation (7) by
$N/N_{\rm p}+ \beta$. The second term, $\beta$, indicates the number of
interaction lists imported from the neighboring domains. It depends on
the opening parameter, $\theta$, and its dependences on $N$ and $N_{\rm
p}$ are very weak. In the case of $N=256^3$, $\theta=1.0$ and $N_{\rm
p}=8$, $\beta \sim 5\times 10^5$.

\section{Conclusion and Discussion}\label{sec:conclusion}

In this paper, we present the PPPM and TreePM implementation of
cosmological N-body simulations on GRAPE-5 and GRAPE-6A for both of a
single processor and parallel systems. The most remarkable feature of
our implementations is that the CPU time per timestep is almost constant
irrespective of the clustering degree of the particles. As for the PPPM
method, the use of GRAPE overcomes its main disadvantage in which PP
calculation cost significantly increases as particle clustering
develops. Under strong particle clustering, our PPPM implementation with
GRAPE-5 and -6A is 4--5 times faster than that without GRAPE, and
sufficiently fast even compared with the PPPM implementation on
MDGRAPE-2 shown in \citet{Susukita2004}. In comparison between the PPPM
and TreePM implementations on GRAPE, the TreePM method has better
performance (i.e. shorter CPU time), although the PPPM method has
advantage in its memory consumption and force accuracy. The CPU time per
timestep in the TreePM method with opening parameter $\theta=1.0$ is
less than half of that in the PPPM method.

Since smaller opening parameter gives better force accuracy but poorer
performance, trade-off between force accuracy and performance has to be
made in the TreePM method. According to
figures~\ref{fig:force_error_N64} and \ref{fig:force_error_N128}, the
force accuracy for $\theta=1.0$ is as good as that of the PPPM method
when the standard PM grid spacing of $H_0$ is adopted. On the other
hand, in the case of the PM grid spacing of $2H_0$, the force accuracy
is degraded under the nearly homogeneous particle distribution, and
thus, we have to reduce $\theta$ to $\simeq 0.5$ in order to reproduce
the same level of force accuracy as for the standard PM grid
spacing. Therefore, the most optimal set of $\theta$ and $H$ is
$\theta=1.0$ and $H=H_0$ or $\theta=0.5$ and $H=2H_0$. The combination
of $\theta=1.0$ and $H=2H_0$ is acceptable only for the situation where
the particle clustering is strong enough (for example, $z<4$ in the LCDM
simulation with $N=128^3$ particles as shown in
figure~\ref{fig:force_error_N128}). In parallel runs, as described in
\S~\ref{sec:performance}, adopting larger PM grid spacing significantly
contributes to reduce the communication among the parallel nodes.

Further improvement of GRAPE system with up-to-date technology would
provide more powerful computing systems for the PPPM and TreePM
methods. As is clear in the breakdown in Table~\ref{tab:cputime_PPPM}
and \ref{tab:cputime_TreePM}, the total performance is limited by the
communication speed between the host computer and GRAPE. For the
communication, GRAPE-5 and 6A uses rather old technology (PCI interface
32bit/33MHz). We are currently developing new versions of GRAPE system
(GRAPE-7 and GRAPE-DR) which adopt new interface technology, such as
PCI-X and PCI Express.  Using this new interface technology, together
with a new GRAPE processor chip, 3-10 times speedup for single node
performance is expected to be obtained.

\bigskip 

We are grateful to Takayuki Saitoh, Atsushi Kawai, Junichiro Makino and
Naoto Miyazaki for their technical advices and helpful discussions.  We
also thank an anonymous referee for useful comments and suggestions. KY
acknowledge support from the Japan Society for the Promotion of
Science. This research was supported in part by the Grants-in-Aid by
Japan Society for the Promotion of Science (14740127, 15$\cdot$10351)
and by the Ministry of Education, Science, Sports, and Culture of Japan
(16684002).

\end{document}